\documentclass[12pt]{article}%
\usepackage{color}
\usepackage{times,graphics,float}
\usepackage{enumerate}
\usepackage{amsfonts,amsthm,amscd}
\usepackage{amsmath}
\usepackage{amssymb}
\usepackage{geometry, color}
\usepackage{setspace}
\usepackage{float}
\usepackage{subfigure}
\usepackage{caption}
\usepackage[utf8]{inputenc}
\usepackage[english]{babel}
\usepackage[nottoc]{tocbibind}
\usepackage{eso-pic,calc,graphicx}
\usepackage{lipsum}
\newtheorem{theorem}{Theorem}[section]

\theoremstyle{definition}
\newtheorem{definition}{Definition}[section]
\newtheorem{remark}{REMARK}[section]
\newtheorem{proposition}{PROPOSITION}[section]
\newcommand{\R}{\mathbb{R}} 

\usepackage{indentfirst}
\usepackage{multirow}
\usepackage{fancyhdr}
\usepackage{csquotes}
\usepackage[numbers]{natbib}

\author{Wong Ka Chun}
\date{September, 2024}
\begin{document}
\begin{spacing}{1.0}

\begin{center}

\LARGE \bfseries  Extrinsic Principal Component Analysis\\[2cm]

\textsc{\large Ka Chun Wong$^1$, Vic Patrangenaru$^1$, Robert L. Paige$^2$, Mihaela Pricop Jeckstadt$^3$}\\[1cm]

\textsc{\small Florida State University$^1$, Missouri S\&T University$^2$,\\Polytechnic University of Bucharest$^3$\\
USA$^{1,2}$ and Romania$^3$}\\[1cm]

\end{center}

\begin{abstract}
One develops a fast computational methodology for principal component analysis on manifolds. Instead of estimating intrinsic principal components on an object space with a Riemannian structure, one embeds the object space in a numerical space, and the resulting chord distance is used.
This method helps us analyzing high, theoretically even infinite dimensional data, from a new perspective.
We define the extrinsic principal sub-manifolds of a random object on a Hilbert manifold embedded in a Hilbert space, and the sample counterparts.
The resulting extrinsic principal components are useful for dimension data reduction. 
For application, one retains a very small number of such extrinsic principal components for a shape of contour data sample, extracted from imaging data.
\\[1cm]

\textbf{Keywords:} statistics on manifolds, extrinsic analysis, PCA, extrinsic mean, Kendall planar shape

\noindent\textbf{MSC2020:}Primary 62R30, 62H25, 62H35.

\end{abstract}
\end{spacing}

\section{Introduction}

{\bf Principal component analysis} (PCA) is a classical tool in multivariate analysis, which plays an important role in dimension reduction.
Recalling the traditional principal component analysis, which first proposed by Pearson (1901)\citep{pearson1901liii}, is a classical dimension reduction method for high dimensional data.
PCA is widely used in multivariate analysis, helping in searching important covariates, visualizing data and so on.
In shape analysis, mean shape and principal component help in extracting the characteristic of the shape space.
D. G. Kendall (1984)\citep{kendall1984shape} ground breaking paper, first considered shapes of planar configuration of {\em $k$ labeled points (landmarks, k-ads)} as points on a {\bf shape space} $\Sigma_2^k$ that turns out to be homeomorphic to the complex projective plane $\mathbb CP^{k-2}.$
Statisticians started building methodology for Kendall shape, including Kent (1992)\citep{mardia1992art}, Ziezold (1994)\citep{ziezold1994mean} and so on.
Huckemann and Ziezold (2006)\citep{huckemann2006principal} proposed a principal component analysis for Riemannian manifolds based on geodesic distance on the intrinsic metric. That was the beginning of intrinsic PCA on manifolds.
Mardia et. al. (2022)\citep{mardia2022principal} advanced research on nested spheres PCA.
For more reference on the subject of PCA on manifolds see \citep{jung2009pca} \citep{jung2012analysis} \citep{jung2011principal}.

On the other hand, there are also discussion on extrinsic mean for shape analysis, or in general, means on manifolds in Patrangenaru and Ellingson(2016)\citep{patrangenaru2016nonparametric}.
In particular, Patrangenaru (1998)\citep{patrangenaru1998asymptotic} introduced the term of Veronese-Whitney (VW) extrinsic mean planar Kendall shape in terms of the VW embedding of $\mathbb C P^{k-2}$ into the space $S(k-1,\mathbb C)$ of selfadjoint $(k-1) \times (k-1)$ matrices introduced by Kent(1992)\citep{mardia1992art}.
In depth results on the asymptotic distribution of this VW mean shape and the resulting bootstrap distribution are given in Bhattacharya and Patrangenaru (2005)\citep{bhattacharya2005large}, Bandulasiri et al. (2009)\citep{bandulasiri2009nonparametric} and Amaral et al. (2010)\citep{amaral2010bootstrap}. Results were extended to infinite dimensional planar shapes of contours in 
Ellingson et al (2013)\citep{ellingson2013nonparametric}, where a discussion on extrinsic mean of a random object on a Hilbert manifold was first considered, and applied to mean Kendall shapes of random contours; as an application, a comparison of the contour of ( the midsection of) the Corpus Callosum (CC) of Albert Einstein with the VW mean CC of senior individuals was given by Qiu et al (2014)\citep{qiu2014far}.

In this paper we propose a method of PCA based on the chord distance on a manifold embedded in an Eucidean space. 
This approach via the PCA in the ambient space where the manifold is embedded has the advantage of being faster and conceptually more consistent that the intrinsic PCA, since from the onset the extrinsic principal submanifolds are going through the extrnsic mean, unlike the intrinsic approach that does not assure this basic compatibility, as shown by Huckeman and Ziezold(2006)\citep{huckemann2006principal}. Our novel approach consists in conducting PCA on the tangent space at the extrinsic mean of the embedded manifold, a method that is a more efficient way than the intrinsic PCA one.

Section 2 will briefly remind the notion of dimensionality for object data on manifold.
In Section 3 we recall basic result on extrinsic means, including definition, uniqueness and computation.
Section 4 is dedicated to introducing extrinsic principal component analysis, and provide some basic related results.
In Section 5, we introduce the reader to Kendall shapes of planar contours. A concrete example of a drastic dimension reduction for Kendall shape of planar contour data extracted from camera images is given here as well. 


\section{Dimensions and manifolds}
In data analysis, the first considerations are about the dimensionality of the data; especially when this dimension is high including in  image analysis, bio-informatics or functional data analysis.
Basically {\em data dimension} is the  {\em local} number of covariates fully describing the data.
Functional data is assumed to be infinitely dimensional, although it is impossible to measure infinitely many covariates.
When it comes to imaging data, widely available to users, it is more difficult to define "dimension", as it depends on many factors, including RGB and relative position of the observer facing the imaged scene.
The issue of image data dimensionality could be solved only by introducing various concepts of {\bf shape}.
For example two planar rigid configurations of points have the same {\bf Kendall shape} if they differ by a direct similarity;
two planar rigid configurations pictured from different remote view points have the same {\bf affine shape}; and,
two planar rigid configurations pictured from different arbitrary view points have the same {\bf projective shape}.
3D Kendall shapes, 3D affine shapes or 3D projective shapes are similarly defined.
Different types of shapes of k-ads (configurations of k labeled landmarks) can be represented on corresponding types of shape spaces.
Such object spaces of k-ads are {\em orbifolds} -quotients of manifolds by certain {\em pseudo-group actions}.
Orbifolds are {\bf manifolds} or, in general, {\em stratified spaces} having a dimension, which is the dimension of the tangent space at a given regular point on the space of orbits.
For example the space of planar Kendall shapes of contours is a Hilbert manifold - $\mathbb C P (\mathbb H)$, the projective space of a complex Hilbert space.
The {dimension of a manifold is the dimension, over the reals, of the linear space, modeling that manifold.
For example the dimension of the round unit sphere $\mathbb S^d = \{x \in \mathbb R^{d+1}, \|x\|=1\}$ is $d,$
the dimension of the planar Kendall shape space of $k$-ads is $2k-4$ and the dimension of the projective shape space of $k$-ads in 3D, is $3k-15$ (see Kendall(1984)\citep{kendall1984shape}).
In manifold statistics, we always consider the distance between objects. Once that is known, we have to define the notion of random object, or random element, according to Fr\'echet(1948)\citep{frechet1948elements}.

\begin{definition}
Assume $(\Omega,\mathcal A, \mathbb P)$ is a probability space and $\mathcal B_{\mathcal M}$ is the Borel $\sigma$-algebra on the manifold
$\mathcal M.$ A random object (r.o.) is a function $X:\Omega \to \mathcal M,$ s.t. $\forall B\in \mathcal B_{\mathcal M}$, $X^{-1}(B)\in \mathcal A.$ The probability measure $Q=\mathbb P_X$ associated with $X$ is defined via $Q(B)=\mathbb P(X^{-1}(B)).$
\end{definition}

There are two main types of distances considered on a manifold $\mathcal M$ (see Patrangenaru and Ellingson (2016)\citep{patrangenaru2016nonparametric}).
One is geodesic distance $\rho^g$ associated with a Riemannian structure $g$ on $\mathcal M.$
The other is chord distance $\rho_j$ associated with an embedding $j :\mathcal M \to \mathbb R^N.$
A statistical data analysis on a manifold is intrinsic, if the distance considered is a geodesic distance, and, it is extrinsic, if the distance considered is a chord distance.
We can see the intrinsic metric, even in simple cases, such as that of a r.o. of a round sphere, leads to iterative algorithms for computing the intrinsic sample mean, so the calculations will be time consuming, cutting in the lifeline of the user.
Most of the time, extrinsic data analysis is faster, since the extrinsic mean is obtained immediately by projecting the mean in the ambient space on the image of the embedded manifold $j(\mathcal M).$  Therefore, whenever one has a choice, it is preferable to work with a {\em chord distance} (see Bhattacharya et al(2012)\citep{bhattacharya2012extrinsic}).


\section{Extrinsic mean and extrinsic covariance matrix}
In this section, we will introduce the notions of extrinsic mean and of extrinsic covariance matrix of a random object on a manifold, related notations and preliminaries.
A general reference for this section is Patrangenaru and Ellingson(2016)\citep{patrangenaru2016nonparametric}.
We will also show how to compute their sample estimates.
To start with, we first focus on extrinsic mean, before we move on to the extrinsic covariance.

Assume $(\mathcal M,\rho)$ is a complete metric space, with a manifold structure and $Q = P_X$ is a probability measure on $\mathcal M$ associated with a random object $X$.
A {\em Fr\'echet mean} is a minimizer of the Fr\'echet function which is the expected square distance from a point to the random object $X$
\begin{equation} \label{Ffun}
F(x) = \int \rho^2(x,y)Q(dy).
\end{equation}
Consider $j:\mathcal M\to\R^N$ is an embedding on $\mathcal M$ to $\R^N$, with the induced chord distance $\rho_j(x,y) = ||j(x)-j(y)||$.
Assume $(\mathcal M, \rho_j)$ is a complete  metric space such that $j(\mathcal M)$ is a close submanifold of $\R^N$.
Then we have the following

\begin{definition} \label{frechet-mean set}
 Let $Q$ be a probability measure on $\mathcal M$ with the chord distance $\rho_j$.
 The set of minimizers of $\mathcal F$ in \eqref{Ffun} is called the {\em extrinsic mean set} of $Q$.
 If the extrinsic mean set has only one point, that point is called the extrinsic mean and it labeled
$\mu_{j,E}(Q),$ or $\mu_E(Q)$ or $\mu_j,$ or $\mu_E.$
\end{definition}

To understand the extrinsic mean, it is important to understand about the embedding $j$.

\begin{definition}\label{nonfocal-point}
Assume $\rho_0$ is the Euclidean distance in $\mathbb R^N.$ A point
$x$ of $\mathbb R^N$ such that there is a {\it unique} point $p$ in
$\mathcal M$ for which $\rho_0(x,j(\mathcal M)) = \rho_0(x, j(p))$ is called
$j$-nonfocal\index{$j$-nonfocal point}. A point which is not $j$-nonfocal is said to be
$j$-focal\index{$j$-focal point}.
\end{definition}

For example, if $j(x) = x$ is the inclusion map, than it is easy to see that the center of the unit sphere $\mathbb S^N = \{ x , ||x|| = 1\}$ is the only focal point of $j$, since $\rho_0(O,j(\mathbb S^N)) = \rho_0(O, j(p))$ $\forall p \in \mathbb S^N$, where $O$ is the origin.

A probability measure $Q$ on $\mathcal M$ induces a probability
measure $j(Q)$ on $\mathbb R^N.$
\begin{definition}\label{nonfocal} A probability measure $Q$ on $\mathcal M$ is
said to be $j$-nonfocal\index{$j$-nonfocal probability measure}  if the mean $\mu$ of $j(Q)$ is a
$j$-nonfocal point.
\end{definition}

Let $\mathcal{F}^c$ is the set of $j$-nonfocal points.
A {\it projection} $P_j : \mathcal{F}^c \to j(\mathcal M)$ is a function $y=P_j(x)$ such that for any $x \in \mathcal{F}^c$, $y$ is the unique, with $\rho_0(x,j(M)) = \rho_0(x, y)$.

\begin{theorem}\label{t:extrinsic-mean}
If $\mu$ is the mean of $j(Q)$ in $\mathbb R^N$.
Then (a)~{\it the extrinsic mean set} is the set of all points $p\in \mathcal M$, with $\rho_0(\mu,j(p)) = \rho_0 (\mu,j(\mathcal M))$ and (b) If $\mu_{j,E}(Q)$ exists then $\mu$ exists and is $j$-nonfocal and $\mu_{j,E}(Q) =j^{-1}(P_j(\mu))$.
\end{theorem}
\begin{theorem} \label{t:extrinsic-generic} The set of  focal points of a submanifold  $\mathcal M$ of
$\mathbb R^N$ that has no flat points (points of zero curvature) with the induced Riemannian structure, is a closed subset of $\mathbb R^N$  of Lebesgue
measure $0$.
\end{theorem}

\begin{definition} \label{extrinsic-sample-mean-def}
Consider an embedding $j:\mathcal M\to \mathbb R^N.$
Assume $(x_1,...,x_n)$ is a sample from  a $j$-nonfocal probability
measure $Q$ on $\mathcal M$, and the function $p\to \frac{1}{n}\sum^n_{r=1}
\|j(p) - j(x_r)\|^2$ has a unique minimizer on $\mathcal M$; this
minimizer is the {\it extrinsic sample mean}\index{extrinsic sample mean as projection of the mean vector}.
\end{definition}
From Theorem \ref{t:extrinsic-mean} the {\it extrinsic sample mean} is
given by
\begin{equation}\label{extrinsic-sample-mean} \overline{x}_E := j^{-1} \left(
P_j(\overline{j(x)})\right)
\end{equation}

\begin{theorem} \label{t:consist-extrinsic-mean}
(Bhattacharya and Patrangenaru(2003)).
Assume $Q$ is a j-nonfocal probability measure on the manifold $\mathcal M$  and $X = \{ X_1, \dots , X_n\}$ are i.i.d.r.o.'s from $Q$.
(a)~If the sample mean $\overline{j(X)}$ is a $j$-nonfocal point, then the extrinsic sample mean is given by $j^{-1}(P_j(\overline{j(X}))).$
(b)~$\overline{X}_E$ is a consistent estimator of $\mu_{j,E}(Q)$.
\end{theorem}

To sum up, we defined the extrinsic mean above, and provide some theorem about the existence of it.
Also we have mentioned about the extrinsic sample mean, which both help us in computing the extrinsic (sample) covariance matrix.
And now, we start on evaluating the extrinsic covariance matrix.

Assume $\mathcal M$ is a $m$ dimensional manifold and $j:M\to\R^N$ is an embedding on $\mathcal M$ such that $j(\mathcal M)$ is closed in $\R^N$.
$Q$ is a $j$-nonfocal probability measure on $\mathcal M$ such that $j(Q)$ has finite moments of order two (or of sufficiently high order as needed).
Assume $(X_1,\dots,X_n)$ are i.i.d. $\mathcal M$-valued random objects with common probability distribution Q.
Recall the extrinsic mean $\mu_E(Q) = \mu_{j,E}(Q)$ of the measure $Q$ on the manifold $\mathcal M$ relative to the embedding $j$ is the Fr\'echet associated with the restriction to $j(\mathcal M)$ of the Euclidian distance in $\R^N$.
Let $\mu$ and $\Sigma$ be the mean and covariance matrix of $j(Q)$ respectively regarded  as a probability measure on $\mathbb R^N$.
Let $\mathcal{F}$ be the set of focal points of $j(\mathcal M),$ and let  $P_j : \mathcal{F}^c \to j(\mathcal M)$ be the projection on $j(\mathcal M)$.
$P_j$ is differentiable at $\mu$ and has the differentiability class of $j(\mathcal M)$ around any nonfocal point.
In order to evaluate the differential $d_\mu P_j$ we consider a special orthonormal frame field that will ease the computations.
Assume $p \rightarrow (f_1(p), \dots ,f_m(p))$ is a  local frame field on an open subset of $\mathcal M$ such that, for each $ p \in M$, $(d_pj(f_1(p)), \dots ,d_pj(f_m(p)))$ are orthonormal vectors in $\mathbb R^N$.
A local frame field $(e_1(y),e_2 (y) ,\dots, e_N(y))$ defined on an open neighborhood $U \subseteq \mathbb R^N$ is {\it adapted to the embedding $j$ } if it is an orthonormal frame field and

\begin{equation} \label{e:adapted-frame}
(e_r(j(p)) = d_pj(f_r(p)), r = 1, \dots, m, \forall p \in j^{-1}(U).
\end{equation}

Let $e_1,e_2 ,\dots, e_N$  be the canonical  basis of $\mathbb R^N$ and assume $(e_1(y),e_2 (y) , \dots, e_N(y))$ is an adapted frame field around $P_j(\mu)=j(\mu_E)$.
Then $d_\mu P_j(e_b) \in T_{P_j(\mu)}j(\mathcal M)$ is a linear combination of  $e_1(P_j(\mu)),e_2(P_j(\mu)), \dots,e_m(P_j(\mu))$:

\begin{equation}
d_\mu P_j (e_b) = \sum_{a = 1}^m \left(d_\mu P_j(e_b)\right) \cdot e_a
\left(P_j(\mu)\right)e_a\left( P_j(\mu)\right), \forall b = 1, \dots, N.
\end{equation}

By the delta method,  $n^{1\over 2} (P_j(\overline{j(X)}) - P_j(\mu))$ converges weakly to a random vector $V$ having a $\mathcal N_N(0, \Sigma_\mu)$ distribution.
Here $\overline{j(X)} = \frac{1}{n}\sum_{i=1}^{n}j(X_i)$ and
\begin{eqnarray}\label{eq:sigma-mu}
{\Sigma}_\mu = \left[ \sum^m_{a=1} d_\mu P_j (e_b) \cdot e_a(P_j
(\mu)) e_a (P_j (\mu)) \right]_{b=1,...,N} \Sigma \nonumber \\
\left[ \sum^m_{a=1} d_\mu P_j (e_b) \cdot e_a (P_j (\mu)) e_a(P_j(\mu))
\right]_{b=1,...,N}^T,
\end{eqnarray}
where $\Sigma$ is the covariance matrix of $j(X_1)$ w.r.t. the canonical basis $e_1, \dots, e_N.$

The asymptotic distribution $\mathcal N_N( 0 , \Sigma_{\mu} )$ is degenerate and can be regarded as a distribution on $T_{P_j(\mu)}j(\mathcal M)$, since the range of $d_\mu P_j$ is a subspace of  $T_{ P_j(\mu)}j(\mathcal M)$.  Note that

$$d_\mu P_j (e_b) \cdot e_a(P_j(\mu)) =0, \quad \hbox{for}\quad
a = m + 1,\dots, N. $$

We provide below a CLT, which applies to an arbitrary embedding, leading topivots and are independent of the chart used.

The tangential component $tan(v)$ of $ v \in \mathbb R^N$ w.r.t. the basis $e_a (P_j
(\mu))\in T_{P_j (\mu)}j(\mathcal M), a = 1,\dots, m$ is given by
\begin{equation}\label{e:tan-comp}
tan(v) = (e_1 (P_j (\mu))^T v \dots e_m (P_j (\mu))^T v )^T.
\end{equation}

Then the random vector $(d_{\mu_{E}}j)^{-1}(n^\frac{1}{2}(P_j(\overline{j(X)})-P_j(\mu)))$ has the following covariance matrix w.r.t. the basis $f_1(\mu_E),\cdots, f_m(\mu_E)$:
\begin{eqnarray}\label{eq:ext-cov}
\Sigma_{j,E} = (e_a(P_j(\mu))^T\Sigma_{\mu}e_b(P_j(\mu)))_{1 \leq a,b \leq m} = \nonumber \\
\left[ \sum d_\mu P_j (e_b) \cdot e_a (P_j(\mu)) \right]_{a=1,...,m} \Sigma \left[ \sum d_\mu P_j (e_b) \cdot e_a(P_j (\mu))\right]_{a = 1,\dots, m}^T .
\end{eqnarray}

\begin{definition}
The matrix $\Sigma_{j,E}$ given by \eqref{eq:ext-cov} is the {\it extrinsic covariance matrix } of the $j$-nonfocal distribution $Q$ ( of $X_1$) w.r.t. the basis $f_1(\mu_E),\dots, f_m(\mu_E)$.
\end{definition}

When $j$ is fixed in a specific context, the subscript $j$ in $\Sigma_{j,E}$ may be omitted .

\begin{remark}\label{r5:7}
In order to find a consistent estimator of $\Sigma_{j,E}$, note that $\overline {j(X)}$ is a consistent estimator of $\mu$, $d_{\overline{j(X)}}P_j$ converges in probability  to  $d_\mu P_j$, and $e_a(P_j(\overline{j(X)}))$  converges in probability to $e_a(P_j(\mu))$ and, further, $$S_{j,n} = n^{-1} \sum (j(X_r) - \overline{j(X)})(j(X_r) - \overline{j(X)})^T$$ is a consistent estimator of $\Sigma$.
It follows that
\begin{eqnarray}\label{E:3.7}
\left[ \sum_{a=1}^m d_{\overline{j(X)}} P_j (e_b) \cdot
e_a(P_j(\overline{j(X)}))e_a(P_j(\overline{j(X)}))\right]S_{j,n} \nonumber \\
\left[ \sum_{a=1}^m d_{\overline{j(X)}} P_j (e_b) \cdot
e_a(P_j(\overline{j(X)}))e_a(P_j(\overline{j(X)}))\right]^T
\end{eqnarray}
is a consistent estimator of  $\Sigma_{\mu}$, and $tan_{P_j(\overline{j(X)})}(v)$ is a consistent estimator of $tan(v).$
\end{remark}

If we take the components of the bilinear form associated with the matrix \eqref{E:3.7} w.r.t. $e_1(P_j(\overline{j(X)})),e_2(P_j(\overline{j(X)})),...,e_m(P_j(\overline{j(X)}))$, we get a consistent estimator of $\Sigma_{j,E}$, called the {\em the sample extrinsic covariance matrix }, given by
\begin{eqnarray}\label{eq:sample-ext-cov}
S_{j,E,n}=\left[\left[ \sum d_{\overline{j(X)}} P_j (e_b) \cdot
e_a(P_j(\overline{j(X)}))\right]_{a=1,...,m}\right]
\cdot S_{j,n} \nonumber \\
\left[\left[\sum d_{\overline{j(X)}} P_j (e_b) \cdot
e_a(P_j(\overline{j(X)}))\right]_{a=1,\dots,m}\right]^T
\end{eqnarray}

\section{Extrinsic principal components}
Principal component analysis seeks a space of lower dimensionality, known as the principal subspace such that the orthogonal projection of the data points onto the subspace maximize the variance of the projected points.
To achieve this goal, we are looking for the eigenvectors of the covariance matrix as the principal components. There we select the largest eigenvalues, since they contribute to most of the variability in the data.
Similarly, the extrinsic principal components a working in a similar way.
But instead of using the covariance matrix of the data set, we use the sample extrinsic covariance matrix.

\begin{definition}
The {\it extrinsic principal components} of the $j$-nonfocal r.o. $X$ on $\mathcal M$ w.r.t. the basis $f_1(\mu_E),\dots, f_m(\mu_E)$ of the tangent space $T_{ \mu_E}\mathcal M$  are  1D submanifolds of $\mathcal M$ going through the extrinsic mean that are obtained by taking the $j$-preimage of the intersection of the affine subspace generated by the eigenvectors $v_i$, $i=1,\dots,m$ of the matrix $\Sigma_{j,E}$ corresponding to the eigenvalues $\lambda_i$, $i=1,\dots,m$ where $\lambda_i$ are listed in their descending order , and by the orthocomplement of the tangent space at the extrinsic mean to $\mathcal M$, with$\mathcal M$.
Here we assume the eigenvalues are simple.
\end{definition}

\begin{remark}
If the extrinsic covariance has an eigenvalue $\lambda$ with multiplicity $k > 1$, we will define in a similar way the extrinsic principal subset of $X$ corresponding to this eigenvalue as follows :
We will take instead the affine subspace generated by the eigenspace of $\lambda$ and by the orthocomplement of the tangent space of $\mu_E$.
\end{remark}

We can consider the principal subspaces generated by the $d_\mu j$ images of the first $k$ eigenvectors of $\Sigma_E$ and by the orthocomplement as an affine subspace of $\R^N$.
This subspace intersects $j(\mathcal M)$ along a subset which is locally sub-manifold whose $j$-preimage is the principal sub-manifold of $\mathcal M$ including the first principal extrinc curves.
We will give some example with different types of data.

The extrinsic sample principal components associated with a random sample $x_1,\dots,x_n$ are defined by considering in the above definition.
The probability measure $Q$ being the empirical $\hat Q_n = \frac{1}{n} \sum_{i=1}^n \delta_{x_i}$.

\subsection{Simulated example : Spherical Data}
The following example shows the extrinsic principal components of a set of spherical data.
In this example we will illustrate the extrinsic principal components in a graphical way by a simple example.
We pseudo-randomly generate 300 points on a two dimensional unit sphere.
To highlight the result easily, the generated data points are concentrated mainly in one direction.
In this case the projection $P_j$ any  point $x \in \R^3$ on the sphere $S^2$, is given by $P_j(x) = \frac{x}{||x||}$.
Results are shown in figure \ref{sphexp1} and table \ref{sphtab}.
The red great circle in Figure \ref{sphexp1} is the first sample extrinsic principal component and the green circle is the second sample extrinsic principal component. The intercept of these two great circles is the extrinsic sample mean.
As we see, the data mainly distributed along the first principal component.
The first extrinsic principal component explains over 87\% of the data in this example.
To look for the data projected onto the first principal component, we first project the data onto the tangent space of the extrinsic sample mean, $T_{P_{\bar{j(X)}}}j(\mathcal M)$, as shown in Figure \ref{Tps}.
We project the data along the tangent space to the first principal component (the vector in red color).
And the re-projected back those data onto the sphere through the origin.
Result shows in Figure \ref{pjpt}.
The projected data (red points) stick on the first principal components on the sphere (the yellow line) along the shortest distance.

\newpage

\begin{figure}
    \centering
    \includegraphics[width=0.85\linewidth]{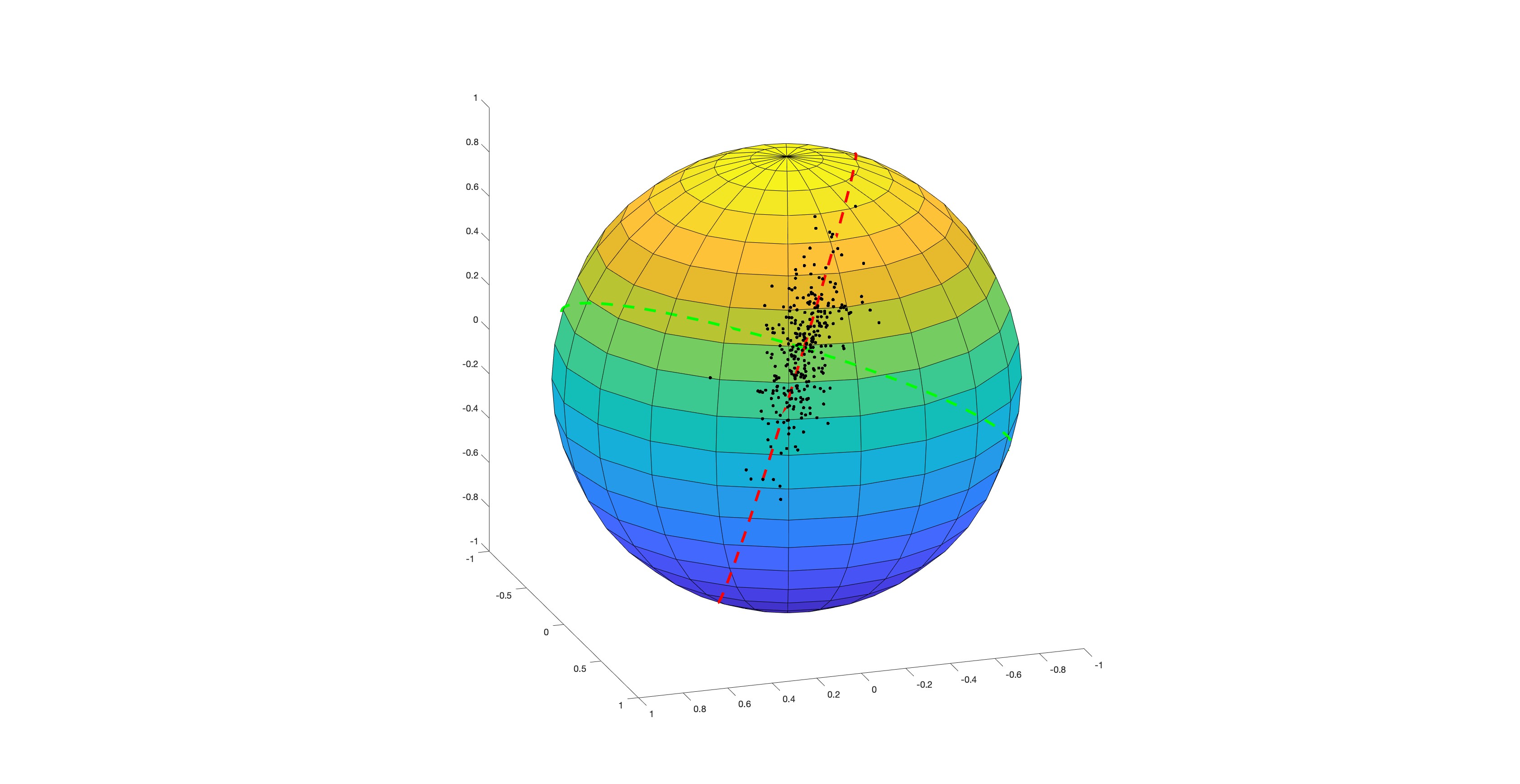}
    \caption{The two principal components in Example 3.1}
    \label{sphexp1}
\end{figure}

\begin{figure}
    \centering
    \includegraphics[width=0.7\linewidth]{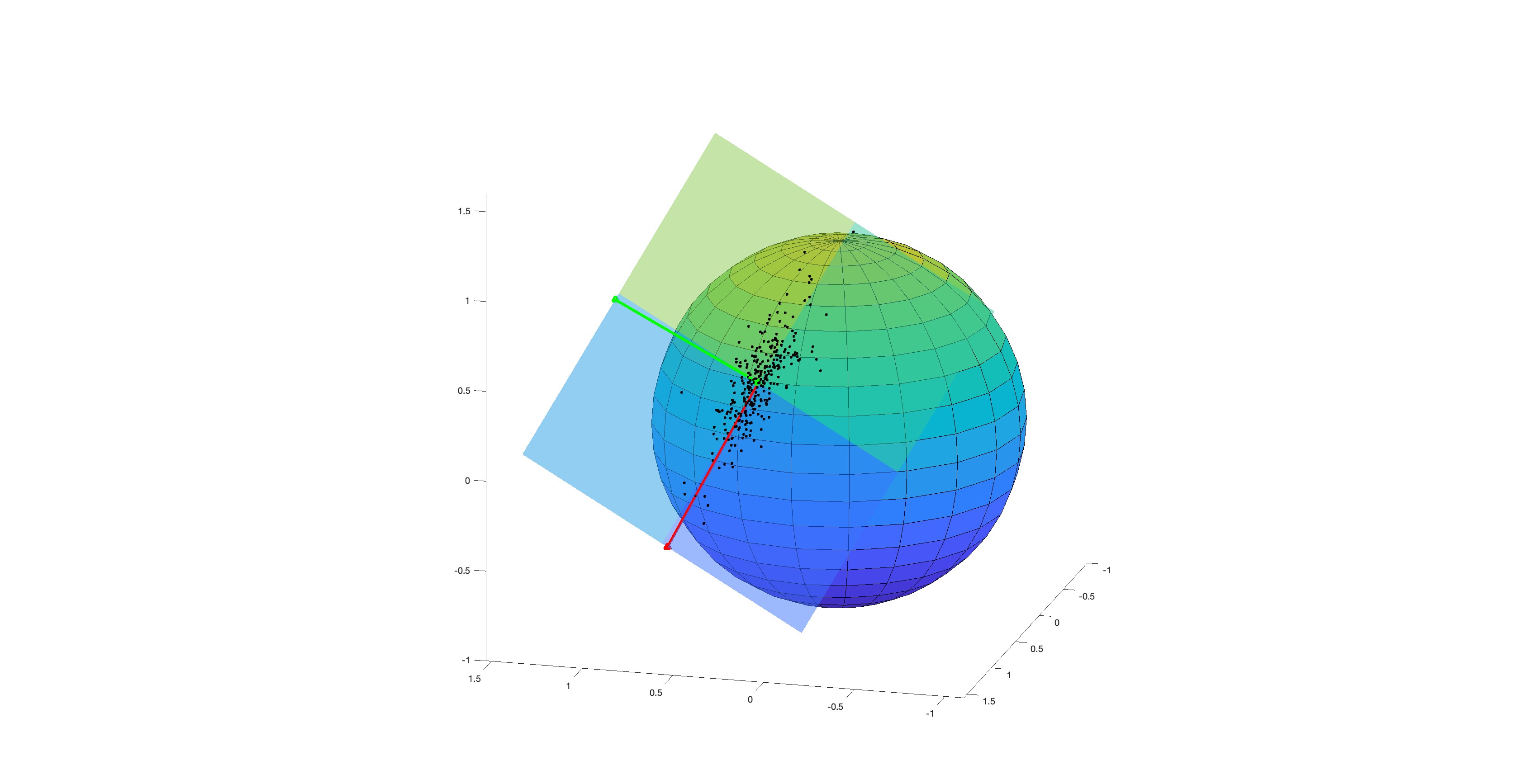}
    \caption{Projected data on $T_{P_{\bar{j(X)}}}j(\mathcal M)$ and The two principal component}
    \label{Tps}
\end{figure}

\begin{figure}
    \centering
    \includegraphics[width=0.7\linewidth]{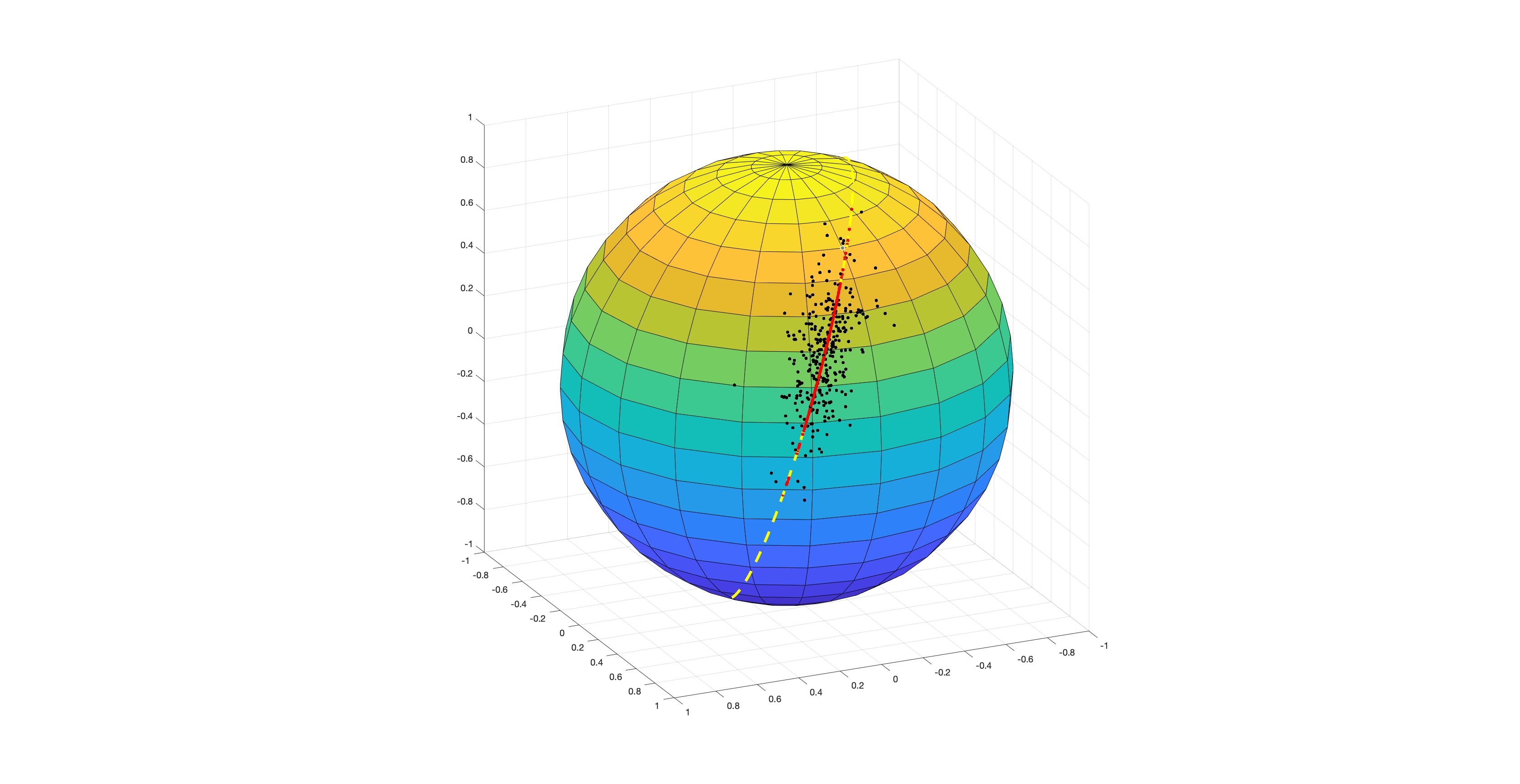}
    \caption{The data projected on the first principal component }
    \label{pjpt}
\end{figure}

\begin{table}[h!]
\centering
\begin{tabular}{ |p{5.5cm}||p{1.5cm} p{1.5cm} p{1.5cm}|  }
 \hline
 Extrinsic sample mean &0.2153 &0.8692 &0.4461\\
 \hline
 \multirow{2}{10cm}{Extrinsic sample covariance} & 0.0045  &-0.0010 & \\
                                                 &-0.0010  & 0.0305  & \\
 \hline
 $f_1(\mu_E)$ &-0.8692 & 0.3775 &-0.3195\\
 \hline
 $f_2(\mu_E)$ &-0.4461 &-0.3195 & 0.8360 \\
 \hline
 1st eigenvalue &0.0305 & &\\
 \hline
 1st p.c. under standard basis &0.4128 & 0.3336 &-0.8475\\
 \hline
 2nd eigenvalue &0.0044 & &\\
 \hline
 2nd p.c. under standard basis &0.8854 &-0.3651 & 0.2876\\
 \hline
\end{tabular}
 \caption{Statistics for example 3.1}
 \label{sphtab}
\end{table}

\section{Extrinsic Principal Components for Shapes of Planar Contours}
In this section we will focus on shape analysis of planar contours.
We will introduce the corresponding shape space and some statistics on this object space.
We will also give an extrinsic principal component analysis concrete example.
The general reference for this section is Patrangenaru and Ellingson(2016)\citep{patrangenaru2016nonparametric}.

Kendall(1984) \citep{kendall1984shape} showed that the space of direct similarity shapes of $k$ planar landmarks can be represented as the manifold $\mathbb C P^{k-2}$.
More general, this is extended here to direct similarity shapes of planar contours.

We focus on contours, boundaries of 2D topological disks in the plane.
To keep the data analysis stable, and to assign a {\em unique} labeling, we make the {\em generic} assumption that across the population there is a unique anatomical or geometrical landmark starting point $p_0$ on such a contour of perimeter one, so that the label of any other point $p$ on the contour is the "counterclockwise" travel time at constant speed from $p_0$ to $p$.
A {\em regular contour} $\tilde \gamma$ is regarded as the range of a piecewise differentiable {\em regular } arclength parameterized function $\gamma: [0, L] \rightarrow \mathbb C, \gamma(0) = \gamma(L),$ that is one-to-one on $[0, L).$
Two contours $\tilde \gamma_1, \tilde \gamma_2$ {\em have the same direct similarity shape} if there is a direct similarity $S : \mathbb C \to \mathbb C,$ such that $S(\tilde \gamma_1) = \tilde \gamma_2$.
Two regular contours $\tilde \gamma_1, \tilde \gamma_2$ have the same similarity shape if their centered counterparts satisfy to $\tilde \gamma_{2,0}=\lambda\tilde \gamma_{1,0},$ for some $\lambda \in \mathbb C \backslash 0$.
Therefore $\Sigma_2^{reg},$ {\em set of all direct similarity shapes of regular contours,} is a dense and open subset of $P({\mathbf H})$, the projective space corresponding to the Hilbert space ${\mathbf H}$ of all square integrable centered functions from $S^1$ to $\mathbb C.$ (see Ellingson et al (2013)\citep{ellingson2013nonparametric}).

The space $P({\mathbf H})$ is a Hilbert manifold.
We here introduce the Veronese-Whitney (VW) embedding $j:P({\mathbf H}) \to \mathcal{L}_{HS}=\mathbf{H} \otimes \mathbf{H}$ given by
\begin{equation}\label{veronese1}
j([\gamma]) = \frac{1}{\|\gamma\|^2}\gamma \otimes \gamma^*, [\gamma] \in P(\mathbf H).
\end{equation}
The {\em Veronese-Whitney mean ( VW mean) } is the extrinsic mean for a random object $X = [\Gamma]$ on $P(\mathbf H)$ with respect to the VW embedding.
The VW extrinsic mean is $[e_1]$, where $e_1$ is the eigenvector corresponding to the largest eigenvalue of $E(\frac{1}{\|\Gamma\|^2}\Gamma \otimes \Gamma^*)$.
The VW extrinsic sample mean can be compute in a similar way.

\begin{proposition}\label{vwmean}
Given any VW-nonfocal probability measure $Q$ on $P(\mathbf H)$, then if $X_1,\dots,X_n$ is a random sample from $\Gamma$, then the VW sample mean $\hat{\mu}_{E,n}$ is the projective point of the eigenvector corresponding to the largest eigenvalue of $\frac{1}{n} \sum^n_{i=1} \frac{1}{\|X_i\|^2}X_i \otimes X_i^*$.
\end{proposition}

Once we have compute the extrinsic mean, the next step is the extrinsic covariance matrix.
The following result of Prentice (1984) \citep{prentice1984distribution} is also needed in the sequel.

\begin{proposition}\label{p10:2}(Prentice (1984) \citep{prentice1984distribution})
Assume $[X_i]$, $\|X_i\|=1$, $i=1,...,n$ is a random sample from a $j$-nonfocal, probability measure $Q$ on $\mathbb{R}P^{N-1}$.
Then the sample (VW-)extrinsic covariance matrix ${S_{j,E}}$ is given by \index{extrinsic covariance}
\begin{equation}\label{eq:prent}
{S_{j,E}}_{ab} = n^{-1} (\eta_N - \eta_a)^{-1} (\eta_N - \eta_b)^{-1} \sum_i (m_a \cdot X_i)(m_b \cdot X_i)(m\cdot X_i)^2,
\end{equation}
where $\eta_a, a =1,...,N,$ are eigenvalues of $K := n^{-1} \sum^n_{i=1} X_i X^t_i$ in increasing order and $m_a, a =1,...,N,$ are corresponding linearly independent unit eigenvectors.
\end{proposition}

Here we give a proof of formula \eqref{eq:prent}.
Since the map $j$ is equivariant, w.l.o.g. one may assume that $j(\overline{X}_E)=P_j(\overline{j(X)})$ is a diagonal matrix, $\overline{X}_E = [m_N]=[e_N]$ and the other unit eigenvectors of $\overline{j(X)}=D$ are $m_a=e_a, \forall a=1,...,N-1$.
We evaluate $d_D P_j$.
Based on this description of $T_{[x]}\mathbb{R}P^{N-1}$, one can select in $T_{P_j(D)}j(\mathbb{R}P^{N-1})$ the orthonormal frame $e_a(P_j(D)) = d_{[e_N]}j(e_a)$. Note that $S(N,\R)$ has the orthobasis $F^b_a, b \leq a$ where, for $a<b$, the matrix $F^b_a$ has all entries zero except for those in the positions $(a,b), (b,a)$ that are equal to $2^{-{1\over 2}};$ also $F^a_a = j([e_a])$.
A straightforward computation shows that if $\eta_a, a =1,...,N,$ are the eigenvalues of $D$ in their increasing order, then $d_D P_j(F^b_a) = 0, \forall b \leq a < N$ and $d_D P_j(F^N_a) = (\eta_N - \eta_a)^{-1} e_a(P_j(D));$ from this equation it follows that, if $\overline{j(X)}$ is a diagonal matrix $D$ then the entry ${S_{j,E}}_{ab}$ is given by
\begin{equation}\label{prent-eq}
{S_{j,E}}_{ab} = n^{-1} (\eta_N - \eta_a)^{-1} (\eta_N- \eta_b)^{-1} \sum_i X_i^a X_i^b (X_i^N)^2.
\end{equation}
Taking $\overline{j(X)}$ to be a diagonal matrix and $ m_a = e_a$ formula \eqref{eq:prent} follows.

\subsection{Data Driven Example}
Here we illustrate an example for extrinsic principal component analysis for planar contour.
Consider the samples of contours of butterfly from Sharvit et al.(1998)\citep{sharvit1998symmetry}.
Some samples are shown in Figure \ref{hand}.
There are 16 contours, each have 500 sampling points.
Each sample contours is a $2 \times 500$ real matrix, each column represent a point of the contour.
We transfer the sample in to a $1 \times 500$ complex matrix, in result the whole sample data will be a $16 \times 500$ complex matrix.
We compute the extrinsic sample mean by using Proposition \ref{vwmean}.
The result mean shape shows in Figure \ref{handmean}; it is smoother than those original sample contours, due to the averaging process. This is always expected when sharp features appear at various locations on individual observations. 
And then we compute the extrinsic sample covariance matrix by using equation (\ref{eq:prent}).
By applying eigenvalue decomposition on the extrinsic sample covariance matrix, we can extract now the extrinsic principal components.
Figure \ref{PCAcont} shows the scree plot for the extrinsic PCA associated with this data set.
The first two sample extrinsic principal component explain almost 90\% of this data set.

\begin{figure}
    \centering
    \includegraphics[width=1\linewidth]{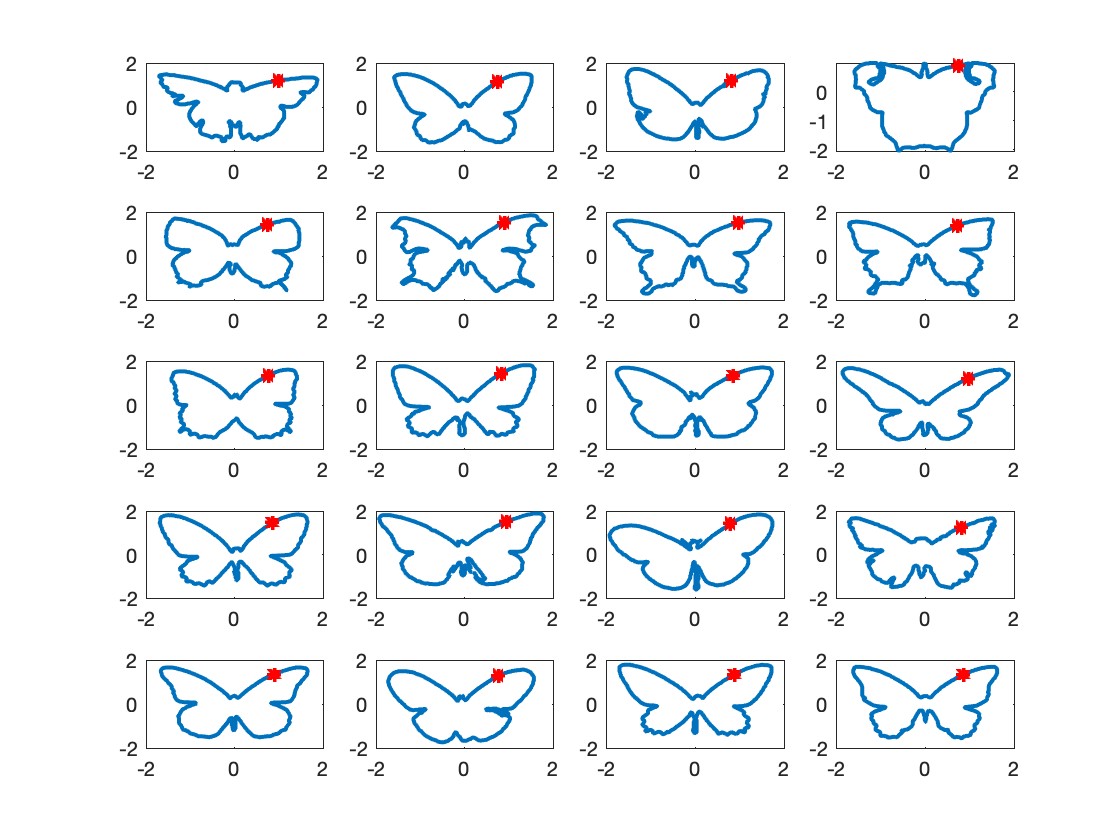}
    \caption{Contours of butterfly}
    \label{hand}
\end{figure}

\begin{figure}
    \centering
    \includegraphics[width=0.5\linewidth]{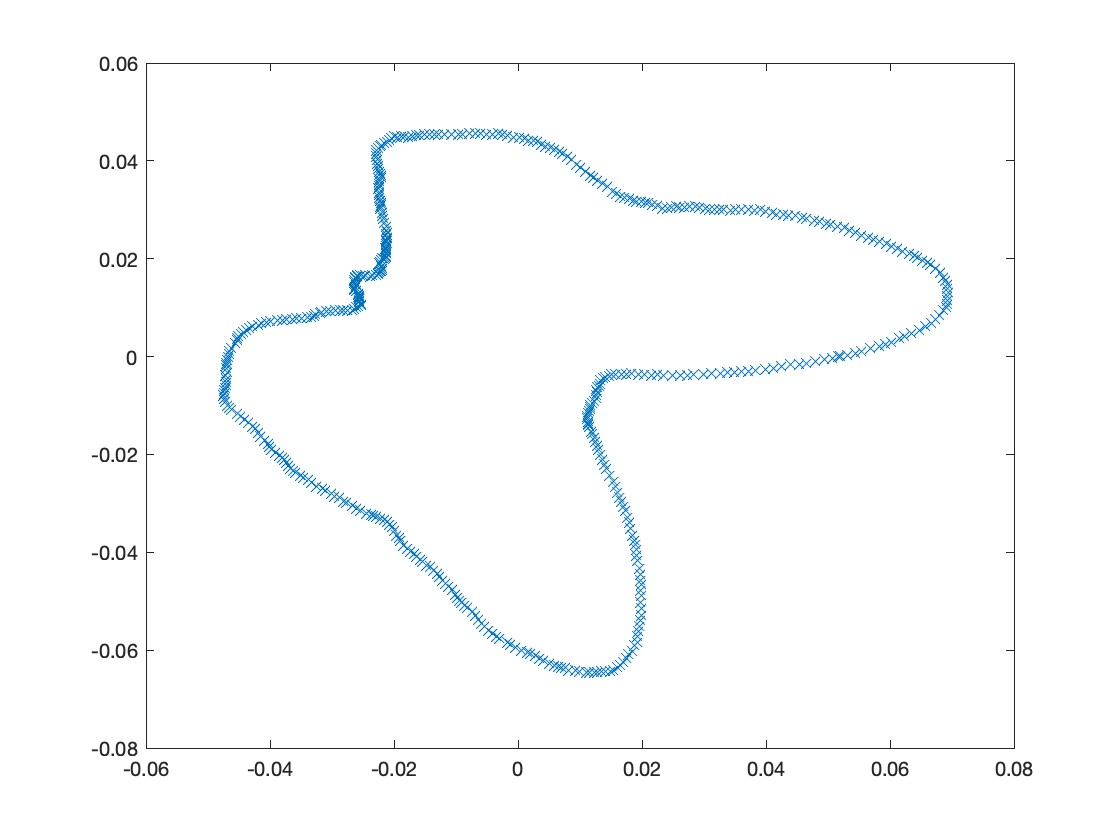}
    \caption{The mean shape of butterffly contours.}
    \label{handmean}
\end{figure}

\begin{figure}
    \centering
    \includegraphics[width=0.5\linewidth]{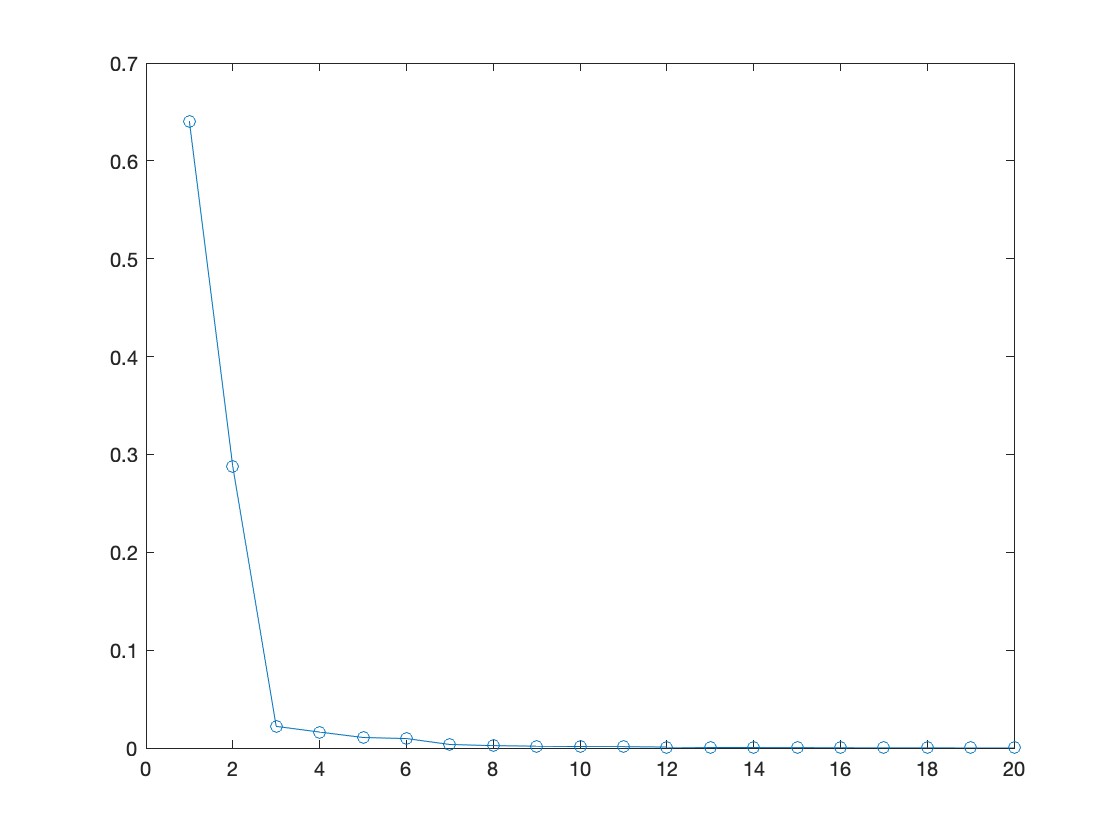}
    \caption{The scree plot for the butterfly data.}
    \label{PCAcont}
\end{figure}

\section{Acknowledges}

Page and Patrangenaru thank the National Science Foundation for awards NSF-DMS:23\\11058, NSF-DMS:2311059.
Pricop Jeckstadt acknowledges support from M-ERA Net Project SMILE, Grant number 315/2022.
She would also like to thank the Isaac Newton Institute for Mathematical Sciences, Cambridge, for support and hospitality during the programme "Discretization and recovery in high-dimensional spaces",where work on this paper was undertaken. 
This work was supported by EPSRC grant EP/R014604/1.

\bibliographystyle{plain}
\bibliography{mybib}

\begin{thebibliography}{10}

\bibitem{amaral2010bootstrap}
Getulio~JA Amaral, Ian~L Dryden, Vic Patrangenaru, and Andrew~TA Wood.
\newblock Bootstrap confidence regions for the planar mean shape.
\newblock {\em Journal of Statistical Planning and Inference},
  140(11):3026--3034, 2010.

\bibitem{bandulasiri2009nonparametric}
Ananda Bandulasiri, Rabi~N Bhattacharya, and Vic Patrangenaru.
\newblock Nonparametric inference for extrinsic means on
  size-and-(reflection)-shape manifolds with applications in medical imaging.
\newblock {\em Journal of Multivariate Analysis}, 100(9):1867--1882, 2009.

\bibitem{bhattacharya2005large}
Rabi Bhattacharya and Vic Patrangenaru.
\newblock Large sample theory of intrinsic and extrinsic sample means on
  manifolds—ii.
\newblock 2005.

\bibitem{bhattacharya2012extrinsic}
Rabindra~N Bhattacharya, L~Ellingson, X~Liu, V~Patrangenaru, and M~Crane.
\newblock Extrinsic analysis on manifolds is computationally faster than
  intrinsic analysis with applications to quality control by machine vision.
\newblock {\em Applied Stochastic Models in Business and Industry},
  28(3):222--235, 2012.

\bibitem{ellingson2013nonparametric}
Leif Ellingson, Vic Patrangenaru, and Frits Ruymgaart.
\newblock Nonparametric estimation of means on hilbert manifolds and extrinsic
  analysis of mean shapes of contours.
\newblock {\em Journal of Multivariate Analysis}, 122:317--333, 2013.

\bibitem{frechet1948elements}
Maurice Fr{\'e}chet.
\newblock Les {\'e}l{\'e}ments al{\'e}atoires de nature quelconque dans un
  espace distanci{\'e}.
\newblock In {\em Annales de l'institut Henri Poincar{\'e}}, volume~10, pages
  215--310, 1948.

\bibitem{huckemann2006principal}
Stephan Huckemann and Herbert Ziezold.
\newblock Principal component analysis for riemannian manifolds, with an
  application to triangular shape spaces.
\newblock {\em Advances in Applied Probability}, 38(2):299--319, 2006.

\bibitem{jung2012analysis}
Sungkyu Jung, Ian~L Dryden, and James~Stephen Marron.
\newblock Analysis of principal nested spheres.
\newblock {\em Biometrika}, 99(3):551--568, 2012.

\bibitem{jung2011principal}
Sungkyu Jung, Mark Foskey, and JS~Marron.
\newblock Principal arc analysis on direct product manifolds.
\newblock 2011.

\bibitem{jung2009pca}
Sungkyu Jung and J~Stephen Marron.
\newblock Pca consistency in high dimension, low sample size context.
\newblock 2009.

\bibitem{kendall1984shape}
David~G Kendall.
\newblock Shape manifolds, procrustean metrics, and complex projective spaces.
\newblock {\em Bulletin of the London mathematical society}, 16(2):81--121,
  1984.

\bibitem{mardia2022principal}
Kanti~V Mardia, Henrik Wiechers, Benjamin Eltzner, and Stephan~F Huckemann.
\newblock Principal component analysis and clustering on manifolds.
\newblock {\em Journal of Multivariate Analysis}, 188:104862, 2022.

\bibitem{mardia1992art}
Kantilal~Varichand Mardia.
\newblock The art of statistical science. a tribute to gs watson.
\newblock 1992.

\bibitem{patrangenaru1998asymptotic}
Victor Patrangenaru.
\newblock {\em Asymptotic statistics on manifolds and their applications}.
\newblock Indiana University, 1998.

\bibitem{patrangenaru2016nonparametric}
Victor Patrangenaru and Leif Ellingson.
\newblock {\em Nonparametric statistics on manifolds and their applications to
  object data analysis}.
\newblock CRC Press, Taylor \& Francis Group Boca Raton, 2016.

\bibitem{pearson1901liii}
Karl Pearson.
\newblock Liii. on lines and planes of closest fit to systems of points in
  space.
\newblock {\em The London, Edinburgh, and Dublin philosophical magazine and
  journal of science}, 2(11):559--572, 1901.

\bibitem{prentice1984distribution}
Michael~J Prentice.
\newblock A distribution-free method of interval estimation for unsigned
  directional data.
\newblock {\em Biometrika}, 71(1):147--154, 1984.

\bibitem{qiu2014far}
Mingfei Qiu, Vic Patrangenaru, and Leif Ellingson.
\newblock How far is the corpus callosum of an average individual from albert
  einstein’s.
\newblock In {\em Proceedings of COMPSTAT-2014, The 21st International
  Conference on Computational Statistics}, pages 403--410. Citeseer, 2014.

\bibitem{sharvit1998symmetry}
Daniel Sharvit, Jacky Chan, H{\"u}seyin Tek, and Benjamin~B Kimia.
\newblock Symmetry-based indexing of image databases.
\newblock {\em Journal of Visual Communication and Image Representation},
  9(4):366--380, 1998.

\bibitem{ziezold1994mean}
Herbert Ziezold.
\newblock Mean figures and mean shapes applied to biological figure and shape
  distributions in the plane.
\newblock {\em Biometrical journal}, 36(4):491--510, 1994.

\end{thebibliography}
\end{document}